\documentclass[pdflatex,sn-nature]{sn-jnl}

\usepackage{graphicx}%
\usepackage{multirow}%
\usepackage{amsmath,amssymb,amsfonts}%
\usepackage{amsthm}%
\usepackage{mathrsfs}%
\usepackage[title]{appendix}%
\usepackage{xcolor}%
\usepackage{textcomp}%
\usepackage{manyfoot}%
\usepackage{booktabs}%
\usepackage{algorithm}%
\usepackage{algorithmicx}%
\usepackage{algpseudocode}%
\usepackage{listings}%
\usepackage{comment}
\usepackage{float}
\usepackage[caption = false]{subfig}
\usepackage{booktabs}
\usepackage{natbib}
\usepackage{geometry}
\usepackage{algpseudocode}
\usepackage{subcaption, caption}
\usepackage{multirow}
\usepackage{import}

\usepackage{hyperref}

\usepackage[margins]{trackchanges}

\usepackage{lineno}

\theoremstyle{thmstyleone}%
\theoremstyle{thmstyletwo}%

\theoremstyle{thmstylethree}%

\newcommand{\beq}{\begin{equation}}
\newcommand{\eeq}{\end{equation}}

\begin{document}

\title[Article Title]{Temporal Complexity and Self-Organization in an Exponential Dense Associative Memory Model}
\date{}

%

\author*[1,2]{\fnm{Marco} \sur{Cafiso}}\email{marco.cafiso@cnr.it}
\equalcont{These authors contributed equally to this work.}
\author*[2,3]{\fnm{Paolo} \sur{Paradisi}}\email{paolo.paradisi@cnr.it}
\equalcont{These authors contributed equally to this work.}
\affil[1]{\orgdiv{Department of Physics 'E. Fermi', University of Pisa, Largo Bruno Pontecorvo 3, I-56127, Pisa, Italy} 
} 
\affil[2]{\orgdiv{Institute of Information Science and Technologies ‘A. Faedo’ (ISTI-CNR), Via G. Moruzzi 1, I-56124 Pisa, Italy}
}
\affil[3]{\orgdiv{BCAM-Basque Center for Applied Mathematics, Alameda de Mazarredo 14, E-48009 Bilbao, Basque Country, Spain} 
}

\abstract{
Dense Associative Memory (DAM) models generalize the classical Hopfield model by incorporating 
$n$-body or exponential interactions that greatly enhance storage capacity. 
%
While the criticality of 
DAM models has been largely investigated, mainly within a statistical equilibrium picture, little attention has been devoted to the temporal self-organizing behavior induced by learning.\\
%
In this work, we investigate the behavior of a stochastic exponential DAM (SEDAM) model through the lens of Temporal Complexity (TC), a framework that characterizes complex systems by intermittent transition events between order and disorder and by scale-free temporal statistics. Transition events associated with birth-death of neural avalanche structures are exploited for the TC analyses and compared with analogous transition events based on coincidence structures.
We systematically explore how TC indicators depend on control parameters, i.e., noise intensity and memory load.\\
Our results reveal that the SEDAM model exhibits regimes of complex intermittency characterized by nontrivial temporal correlations and scale-free behavior, indicating the spontaneous emergence of self-organizing dynamics. 
Notably, such regimes arise over finite intervals of noise intensity rather than at a single critical point, consistent with the concept of \textit{extended criticality}. Further, the noise intensity range needed to reach this critical region, where self-organizing behavior emerges, slightly decreases as the memory load increases.
This study highlights the relevance of TC as a complementary framework for understanding learning and information processing in artificial and biological neural systems, revealing the link between the memory load and the self-organizing capacity of the network.
}






\maketitle

\section{Introduction}
\label{intro}

\noindent
The Hopfield network (HN) \cite{hopfield_pnas1982}, as such as other similar models \cite{little_mb1978,amari_bc1978},is a well-known neural model with associative memory implemented through a Hebbian learning rule \cite{hebb_1949,brownARN1990_HebbianReview}.
%
In a nutshell, this rule states that ``neurons that fire together, wire together'' \cite{Shatz1992DevelopingBrain}.
The classical HN is a discrete system in both time and state domains, composed of binary McCulloch and Pitts neurons, usually assuming $+1$ (firing) and $-1$ (no firing) values\footnote{
Alternatively, some authors assume the ``no firing" state labeled by the $0$ value.
}.
In his seminal paper \cite{hopfield_pnas1982}, Hopfield proved that the Hebbian learning rule defines an energy landscape in which each stored pattern corresponds to an energy minimum, that is, a stable equilibrium point in the network dynamics.
%
%
Moreover, after a pattern $\mbox{\boldmath$\xi$}$ is stored in the network, if a distorted version of this pattern $\widetilde{\mbox{\boldmath$\xi$}}$ is presented to the model as its initial state $\mbox{\boldmath$S$}_{t=0}$, the network can ``associate'' this same distorted pattern to the nearest stored one\footnote{
The distorted input must belong to the attraction basin of the stored pattern. Further, memory association is defined as successful within a small error, such as the fraction of incorrect pixels in an image.
}.
The term ``associative memory'' is actually used to underline this capacity of the HN to recall the original stored pattern, or memory, $\mbox{\boldmath$\xi$}$ from a initial noisy or incomplete version of the same pattern $\widetilde{\mbox{\boldmath$\xi$}}$.
%
%
%
It has been widely demonstrated that the 
retrieval process remains effective as long as the number of stored patterns $K$, also referred to as the {\it network memory load}, stays below a critical threshold, thus identifying a {\it retrieval phase} \cite{amit_pra1985,forrest_jpa1988}.
Beyond this threshold a critical transition occurs between the retrieval phase and a sort of ``forgetting phase''. This is a condition, also named {\it catastrophic forgetting} \cite{amit_ap1987}, in which the system is unable to correctly retrieve the stored patterns and the HN model loses its associative memory ability. 

\noindent
To extend the network loading capacity, in the sense of maintaining a high accuracy in the pattern retrieval,
many network model extending the classical HN have been proposed considering generalizations of the Hebbian rule. In particular the n-body Dense Associative Memory (DAM) models were proposed and investigated since the 1980' \cite{gardner_npsb1985,lee_pd1986,peretto_bc1986,gardner_epl1987,gardner_jpamg1987} culminating nowadays in a broad array of studies \cite{krotov2016modernHopf,krotov_pnas2019,agliari_epjp2020,krotov_nrp2023,agliari-pedreschi_pa2023_unsupervised,agliari-pedreschi_pa2023_unsupervised,lucibello_prl2024,theriault_spp2024,serricchio_nn2025}.
%
%
%
%
While the classical HN includes only pairwise (2-body) interactions, the DAM model incorporates higher-order, $n$-body interactions. 
Further extensions of this model based on an exponential interaction function have been proposed and investigated 
\cite{demircigil2017modernHopfExp,albanese_pa2026}.

\noindent
The criticality of n-body DAM models has been also largely investigated since the 1980' \cite{gardner_npsb1985,gardner_epl1987,gardner_jpamg1987,amit_pra1985,amit_ap1987,amit_book2005_critical_statphys,krotov2016modernHopf}, while only a few recent works have so far explored the criticality arising in the {\it exponential DAM} model \cite{ramsauer2021hopfieldnetworksneed,lucibello_prl2024, cafiso2025criticalitystochasticmodernhopfield}, i.e., the DAM model with exponential interaction in the energy function \cite{demircigil2017modernHopfExp}. 
%
%
Further, the investigation of critical behavior in Hopfield-type associative memory models has been grounded on the equilibrium 
thermodynamics framework of canonical ensemble \cite{huang_1987_critical_statphys}, thus considering the neural network in thermal equilibrium with a thermal bath at given temperature $T > 0$ \cite{amit_pra1985}\footnote{
 This assumption naturally follows from the similarity of the HN with the Ising spin model at temperature $T=0$ \cite{Ising1925OriginalIsingModel}
}.
%
%

\noindent
Conversely, to the best of our knowledge, DAM models have not yet been studied from the perspective of complexity, which relies on the emergence of self-organization in cooperative dynamics, and, in particular,  through the lens of Temporal Complexity (TC), which primarily investigates
the temporal evolution of self-organizing structures 
\cite {paradisi_csf15_preface,grigolini_csf15_bio_temp_complex,paradisi_springer2017}.  In fact, until now, only few theoretical works, mainly speculative in nature, deal with the complexity of machine learning and artificial intelligence models
\cite{krakauer_fcs2023,danovski_fcs2024}.
%
As it will be detailed in the next section, 
the TC framework basically concerns the intermittent behavior of self-organization, and the {\it complex intermittency} condition is mainly identified by self-similarity, scale-free and, thus, power-law behavior in statistical features 
\cite{grigolini_csf15_bio_temp_complex,paradisi_springer2017}.
The TC property of complex intermittency
was found in both theoretical models \cite{turalska-grigolini_pre11,beig-grigolini_pre15,grigolini_csf15_bio_temp_complex,akin_pa06} and experimental observations, e.g., brain data \cite{,allegrini_fp2010,allegrini_pre10,tagliazucchi-chialvo_fp2012,allegrini_csf13,tagliazucchi-chialvo_fn2016}, biological systems 
\cite{grigolini_csf15_bio_temp_complex,grigolini2021CriticalityCelltoCellCommunication}, nanocrystals \cite{paradisi_aipcp05,bianco_jcp05} and fluid turbulence
\cite{paradisi_chapter2023,paradisi_npg12,paradisi_epjst09}.
%

\noindent
However, while relatively few studies are devoted to TC in large multi-component systems, criticality attracted the interest of a broad scientific community since decades, which can be traced back to the field of theoretical statistical physics
\cite{parisi_prl1979,huang_1987_critical_statphys,amit_book2005_critical_statphys,cardy_2015_critical_statphys}. Over the past two decades, criticality has also found application in various contexts, e.g., neural \cite{beggsplenz_jn2003,beggsplenz_jn2004, allegrini_fp2010,paradisi_aipcp13,allegrini_csf13,fingelkurts_csf2013,allegrini_pre15,Deco2012CriticalityMultistabilityBrainRest,scarpetta2013NeuralAvalanchesatCriticalPoint,sorrentino_ni2023avalanchesbrainfingerprint}, biological  \cite{grigolini_csf15_bio_temp_complex,grigolini2021CriticalityCelltoCellCommunication} and complex systems \cite{turalska-grigolini_pre11,beig-grigolini_pre15,mahmoodi_c2018}, as well as in artificial intelligence (AI) \cite{berstschinger2004RealTimeComputationCriticalRNN, boedecker2012InformationProcessingEchoStateNetworks,schoenholz2017deepinformationpropagation, shriki2016OptimalInformationRepresentationCriticalityRNN}.
In particular, with respect to AI, criticality is expected to play a pivotal role in learning biological processes due to maximization of information transmission in the brain at the critical state 
\cite{maass2002RealTimeComputingWithoutStableStates,hadelman2005CriticalityInRealBrainLearning,west-grigolini_pr2008_fract_events,fraiman_2009_critical_brain,chialvo2010_critical_brain,vanni-grigolini_prl2011_critical,schoenholz2017deepinformationpropagation,Angiolelli2025CriticalityExplainStructureFunctionRelationshipBrain}.

\noindent
In a nutshell, while complexity is meant to be the capacity of triggering emergence of self-organization, criticality involves a sharp transition in the parametric space between different phases.
However, TC is defined by sequences of fast transitions between order and disorder, a concept that recalls that of criticality, since the latter also involves the emergence of an intermediate condition between order and disorder.
The relationship between complexity, in particular temporal complexity, and criticality has been studied in some specific cases \cite{contoyiannis_pla2000,contoyiannis_prl2002,contoyiannis_pre2007,allegrini_fp2010,allegrini_csf13,allegrini_pre15,turalska-grigolini_pre11,beig-grigolini_pre15,grigolini_csf15_bio_temp_complex,tagliazucchi-chialvo_fp2012,tagliazucchi-chialvo_fn2016,zare-grigolini_csf2013,grigolini_chapt2014,mafahim-grigolini_njp2015}, but the general interrelation between these two concepts is still unclear and further in-depth exploration is needed.
%
%

\vspace{.1cm}
\noindent
The working hypothesis of the present work is that the ability of DAM models to accurately recall patterns 
should imply some form of organizing principle in the neural network, thus recalling the basic idea of {\it complexity}, which assumes the spontaneous emergence of self-organizing structures and cooperative dynamics in multi-component systems \cite{sornette_book2006,grigolini_csf15_bio_temp_complex,paradisi_springer2017}. 
Then, the questions that we deal with in this paper are: 
does a DAM model display a complex self-organizing behavior? In more detail, does a DAM model exhibit a condition of complex intermittency in accordance with the TC framework? And, if it does, are there specific regions in the parameter space where TC behavior emerges?
%
%

\noindent
Then, in order to clarify the above questions, in this work we investigate the TC features of 
a stochastic exponential DAM (SEDAM) model with a multiplicative dichotomous noise term, thus extending the model of Demircigil et al. \cite{demircigil2017modernHopfExp}.
This allows us to carry out the study without resorting to the canonical equilibrium ensemble framework, but instead computing and analyzing temporal dynamics without any assumption of steady-state equilibrium. More specifically, the thermodynamical approach to associative memory models does not account for out-of-equilibrium dynamics, which, on the contrary, is the focus of the present work.
Through numerical simulations, we show the behavior of TC indices 
as the noise intensity and the number of stored patterns vary. In particular, we identified three regions: a sub- and super-critical region with a Markovian dynamics, and a critical region marked by the emergence of self-organization.

\vspace{.1cm}
\noindent
The paper is organized as follows. In Section \ref{dam}, we introduce our stochastic version of the DAM model with exponential interaction function. In Section \ref{tc-idc}, after a brief overview of temporal complexity, we introduce the event-based TC analysis applied here. In Section \ref{results} we show and discuss the results of numerical simulations and TC analyses and, finally, in Section \ref{conclusions} we draw some conclusions.

\section{Methods}
\label{sec:Methods}

\subsection{The modern Hopfield neural network model}
\label{dam}

\noindent
Given a network with $N$ neurons and a set of $K$ binary patterns
$\mbox{\boldmath$\xi$}_\mu = [\xi_\mu^1, \xi_\mu^2, \ldots, \xi_\mu^N]$, $1\le \mu \le K$, with $\xi_\mu^i \in \{-1, +1\}$, the Hebbian rule first proposed by Hopfield \cite{hopfield_pnas1982} is given by the following mathematical formula for the {\it weight} or {\it connectivity matrix} ${\bf W}$:
\begin{equation}   
W^{ij} = \frac{1}{N} \sum_{\mu =1}^K \xi_\mu^i \xi_\mu^j\ .
\label{hebb_rule_1}
\end{equation}
Denoting by ${\bf S}_t = [S^1_t, ..., S^N_t]$  the overall network state at discrete time $t$, with $S^i_t \in \{-1, +1\}$ the dichotomous state of the $l$-th neuron, the energy function is then defined by:
\beq
E[\mbox{\boldmath$S$}] = 
- \sum_{
\substack{i,j=1 \\ i\ne j}
}^N W^{ij} S^i_t S^j_t\ .
\label{energy_hn}
\eeq
This formula can be rewritten in the following way \cite{krotov2016modernHopf}:
\begin{equation}
E = -\sum_{\mu=1}^K F(\mbox{\boldmath$\xi$}_\mu^T \mbox{\boldmath$S$}_t)
\label{energy_dam}
\end{equation}
where $F(y) = y^2$ is the interaction function and $\mbox{\boldmath$\xi$}_\mu^T {\bf S}_t$ the scalar product of vectors $\mbox{\boldmath$\xi$}_\mu^T$ and ${\bf S}_t$.
Given the energy function, the dynamical rule of the classical HN is retrieved by applying the generalized energy-based update formula given by  Krotov $\&$ Hopfield \cite{krotov2016modernHopf}:
\begin{equation}
S^i_{t} = \text{sgn} \Big[-E(\mbox{\boldmath$S$}_{t-1}^{(i+)}) + E(\mbox{\boldmath$S$}_{t-1}^{(i-)}) \Big] \quad ;\quad i = 1,...,N \ ,
\label{eqs_dam}
\end{equation}
where $S^i_{t}$ is the state of the $i$-th neuron at time $t$\footnote{
In the first Hopfield formulation \cite{hopfield_pnas1982} and in some later studies, updates are asynchronous, meaning that neurons update at a random time within a given time window. Here we assume a synchronous single neuron dynamics as described by Eq.~(\ref{eqs_dam}). 
}, sgn$[\cdot]$ is the sign function, \(\mbox{\boldmath$S$}_{t-1}^{(i+)}=\mbox{\boldmath$S$}_{t-1}(S^i_{t-1}=1)\) and \(\mbox{\boldmath$S$}_{t-1}^{(i-)}=\mbox{\boldmath$S$}_{t-1}(S^i_{t-1}=-1)\)\footnote{
This dynamical rule assumes a zero firing threshold, but a non-zero threshold $Th$ can be considered by adding a term $-Th$ inside the sign function.
}.
Interestingly, this alternative form of energy and dynamical functions allows straightforward generalizations by considering different shapes of the interaction function $F(y)$. 
In particular, for a monomial, or power-law, of order $n$, i.e., $F(y) = y^n$, we get the DAM model with $n$-body interactions \cite{gardner_npsb1985,krotov2016modernHopf}.

\noindent
Here, we are interested in the exponential DAM model of Demircigil et al. \cite{demircigil2017modernHopfExp}, which assumes an exponential interaction function: $F(y) = e^y$. The set of dynamical equations of the original exponential DAM model is then given by:
\begin{equation}
S^i_t = \text{sgn} \Big[-\sum_{\mu=1}^K \exp(\mbox{\boldmath$\xi$}_\mu^T \mbox{\boldmath$S$}^{(i+)}_{t-1}) + \sum_{\mu=1}^K \exp(\mbox{\boldmath$\xi$}_\mu^T \mbox{\boldmath$S$}^{(i-)}_{t-1}) \Big]
\ ; \quad i = 1, ..., N \ .
\label{eqs_exp_dam}
\end{equation}
Interestingly, the exponential DAM model results in being compatible with the limit, for $n\rightarrow \infty$, of the n-body DAM model \cite{demircigil2017modernHopfExp}.\\
As already said above, the Hopfield and Hopfield-type network models admit an energy landscape with a set of asymptotically stable points with energy minima that, in the case of the exponential DAM, are particularly deep and stable. This implies a very fast convergence through an energy minimum, i.e., a stored pattern, so that the dynamical evolution remains trivial.
However, the presence of sufficiently high noise fluctuations can profoundly alter this stable condition.
To this goal, we generalize the original exponential DAM by introducing a multiplicative noise in the dynamical equations:
\begin{equation}
S^i_t = \epsilon[i,t] \cdot
\text{sgn} \left[-\sum_{\mu=1}^K \exp\left(\mbox{\boldmath$\xi$}_\mu^T \mbox{\boldmath$S$}^{(i+)}_{t-1} \right) + \sum_{\mu=1}^K \exp\left(\mbox{\boldmath$\xi$}_\mu^T \mbox{\boldmath$S$}^{(i-)}_{t-1} \right) \right]
\ ;\quad i = 1, ..., N\ ,
\label{our_model}
\end{equation}
where $\epsilon[i,t]$ is a dichotomous random variable with values in $\{-1, 1\}$ and probabilities $P(\epsilon[i,t] = -1) = p$ and $P(\epsilon[i,t] = +1) = 1 - p$, $0\le p \le 1$.
The stochastic exponential DAM (SEDAM), given in the above Eq. (\ref{our_model}), is the dynamical model being considered henceforth to investigate the role of the number of stored patterns $K$ and of noise intensity, given by the probability $p$ of changing state.
It is worth noting that the SEDAM dynamics given by Eq. (\ref{our_model}) yield a route within the energy landscape that is formally defined by:
\begin{equation}
E_t = E[\mbox{\boldmath$S$}_t] = -\sum_{\mu=1}^K exp(\mbox{\boldmath$\xi$}_\mu^T \mbox{\boldmath$S$}_t)
\label{energy_exp_dam}
\end{equation}

\subsection{Intermittency-Driven Complex Systems and Temporal Complexity Analysis}
\label{tc-idc}

\noindent
A complex system consists of many interacting elements, or nodes, with nonlinear, cooperative dynamics that give rise to emergent, self-organized structures, typically displaying long-range correlations among different elements and in time \cite{grigolini_csf15_bio_temp_complex,paradisi_springer2017}.
%
%
%
These systems are often represented as networks with intricate link structures.
In this approach, the features of a complex network, including those related to the large-scale structures that we are interested in, are defined and evaluated through the lens of topological properties, such as degree distribution and clustering \cite{albert_rmp02,newman_siamr03,boccaletti_pr2006,barrat2008,rubinov-sporns_ni2010,mora_jsp2011,boccaletti_pr2014,paradisi_csf15_preface}.
%
%
However, the temporal patterns of a complex dynamical system are just as important as the topological properties of the network, both at the level of individual nodes and of specific node groups, up to the global level of the network (large-scale dynamical structures) \cite{paradisi_csf15_preface,grigolini_csf15_bio_temp_complex,turalska-grigolini_pre11,paradisi_springer2017}.
In particular, within complex dynamical systems, the class of intermittent systems is particularly relevant and prevalent. In fact, intermittency represents the manifestation of cooperative dynamics through the presence of a large number of regions in the state space with equilibrium saddle points and associated stable and unstable varieties, where in the first case a self-organized metastable state typically emerges while in the second case a rapid transition occurs. 
In this conceptual framework, self-organized structures are 
metastable states with a high level of internal coherence and
a significant lifetime, at the end of which the instability of nonlinear interactions causes a rapid decay of the self-organized state itself and a ideally instantaneous transition to a new state\footnote{
The lifetime is typically much longer that the time scale of the rapid transition to a new state, whatever self-organized or disordered.
}.
%
Each rapid, ideally instantaneous, transition, hereafter named Rapid Transition Event (RTE), is towards another saddle point and, then, another metastable state, thus giving rise to a temporal sequence of crucial transition events.
RTEs are then associated with sudden drops in the system's coherence and memory.
This emergence of RTEs and metastable states has been referred to as Temporal Complexity (TC) \cite{turalska-grigolini_pre11,grigolini_csf15_bio_temp_complex}  or Intermittency-Driven Complexity (IDC) \cite{paradisi_springer2017}. 
A prototypical example of TC/IDC is given by the heteroclinic channel, 
a subset of the configuration space including a sequence of saddle points towards which the systems travel essentially ``jumping'' from a metastable state to another one
\cite{rabinovich_prl2006,rabinovich_pcb2008,rabinovich_plr2012}.
Another reference model for TC/IDC is the Pomeau-Manneville map, which is a well-known prototype of 
type-I intermittency and was widely investigated by Manneville in the 1980' \cite{manneville_jpp1980,pomeau-manneville_cmp1980}
as a model of turbulent bursting. In this case the random transitions
are related to the presence of a marginally stable point and  a ``laminar'' region, associated with long-range correlated motion, and a 
``chaotic'' region with turbulent bursting and memory drop.

\noindent
The central feature of complex systems belonging to the TC/IDC class is therefore a birth-death process of intermittent self-organization that is formally defined by a sequence of crucial transition events, i.e., a birth-death stochastic point process.

\subsubsection{Event-based analysis of Temporal Complexity}

\noindent
The analysis of a TC/IDC complex system then involves the characterization of the birth-death intermittent process of self-organization and, in particular, the statistical analysis of the related sequence of inter-event times (IETs), such as their statistical distribution and their statistical inter-relationship, also including IET-to-IET correlations.\footnote{
The crucial events and their associated IETs can also be labelled in different ways such as, e.g., events triggering coherent, self-organized, states and events determining a switch to a chaotic or random state.
}
\noindent
The birth-death intermittent process 
is mathematically defined as a stochastic point process \cite{cox_1970_renewal,cox_1980_point}, i.e.,
a set, or sequence, of increasing random times, so that the IETs $\tau_n$ are computed as successive event time differences:
\beq
\{t_n\}_{_{n=0}}^{^T}
;\quad t_0=0;\quad t_{n+1} > t_n\ \forall n\ \Rightarrow
\left\{ \tau_n=t_{n}-t_{n-1} \right\}_{_{n=1}}^{^T}
\label{iet}
\eeq
with $t_n$ labelling the $n$-th event occurrence time.
A ubiquitous feature assessing and quantifying system's temporal complexity is the emergence of a slow power-law decay in the tail of the Probability Density Function (PDF) of IETs (IET-PDF) \cite{grigolini_csf15_bio_temp_complex,paradisi_springer2017}:
\beq
\psi(\tau) \sim \frac{1}{\tau^\mu} \ ; \quad \psi(\tau)d\tau = Prob\{\tau \leq T < \tau + d\tau\}d\tau\ .
\label{powerlaw_iet}
\eeq
The underlying point process is then non-Poisson and the system displays long-range correlations and non-Markovian dynamics.
In this case, $\mu$ is exploited as a measure of self-organizing behavior and denoted as TC (or IDC) index or, generally speaking, {\it complexity index}.
Conversely, the emergence of an exponential decay in the IET-PDF, which is the signature of an underlying Poisson process, is associated with non-complex behavior, that is, with short-range correlations and pure random noise.
Statistical interdependence among IETs is also a key property, and, in particular, the system is defined to be a {\it renewal} (point) process when the IETs are mutually independent and identically distributed (i.i.d.) random variables. 
%
Power-law decay and renewal condition are the basic assumptions
for the ``ideal'' intermittency-driven complex systems in
the TC framework. The search for the above conditions is the objective of the Event-Driven Diffusion Scaling (EDDiS) analysis \cite{grigolini2001asymmetric,paradisi_romp12,paradisi_csf15_pandora,paradisi_springer2017,paradisi_chapter2023}, whose application nevertheless goes beyond the specific renewal assumption.
%
%
For reader's convenience and for the sake of completeness, we report in Appendix \ref{app:eddis} some details of the specific EDDiS tool applied here, while the most complete version of the method can be found in Refs. \cite{paradisi_romp12,paradisi_springer2017}. 
Here we just recall that the method is based on deriving different diffusion processes driven the event sequence ${t_n}$ and, then, in estimating some scaling indices of the derived diffusion processes \cite{allegrini_pre10,paradisi_csf15_pandora,paradisi_springer2017}. 
In the present application, 
we limit to estimate the scaling exponents of the so-called Asymmetric Jump (AJ) diffusion process $X_{\rm AJ}(t)$ in the discrete time $t$ (see Appendix \ref{app:eddis}).
This choice is linked to the greater robustness of the AJ rule and its ability to cover a wider range of different scaling behaviours \cite{grigolini2001asymmetric}. In particular, given $\Delta X_{\rm AJ}(t,\Delta t) = X_{\rm AJ}(t+\Delta t) -X_{\rm AJ}(t)$, the second moment scaling $H$ is computed through the 
Detrended Fluctuation Analysis (DFA) (see Appendix \ref{app:dfa} for details):
\beq
{\rm DFA}^2(\Delta t) = \langle  \Delta X^2_{\rm AJ}(t,\Delta t) \rangle_t \sim \Delta t^{2H}\ ,
\label{h_dfa}
\eeq
being $\langle \cdot \rangle_t$ a proper time average also involving a proper detrending function (see Appendix \ref{app:eddis} for details). Similarly, the self-similarity index $\delta$ of the diffusion PDF is computed through the Diffusion Entropy (DE) analysis (see Appendix \ref{app:de} for details):
\begin{eqnarray}    
&&p_{\rm AJ}\left(x,\Delta t\right) dx =   \frac{1}{\Delta t^\delta} F\left( \frac{x}{\Delta t^\delta}  \right) dx\ , \label{delta_dfa} \ ,\\
\ \nonumber  \\
&&p_{\rm AJ}\left(x, \Delta t \right) dx = 
Prob\{  x\le \Delta X_{\rm AJ}(t,\Delta t) < x+dx \}
\ , \nonumber
\end{eqnarray}
The values $H=0.5$ and $\delta=0.5$ denote scaling compatible with normal diffusion, which is the signature of short-range correlations and Markovian non-cooperative dynamics.
Conversely, values $H \ne 0.5$ and $\delta \ne 0.5$ follow from long-range correlations and non-Markovian dynamics, thus marking the presence of self-organizing behavior that was proved to be linked to the emergence of a critical region 
\cite{contoyiannis_pla2000,contoyiannis_prl2002,contoyiannis_pre2007,turalska-grigolini_pre11,beig-grigolini_pre15,zare-grigolini_csf2013,mafahim-grigolini_njp2015}.

\noindent
We also evaluate the interdependence among IETs by computing the usual autocorrelation function, but taking as time the event occurrence index $n$:
\beq
C(\Delta n) 
= \frac{1}{\sigma^2_{\tau}}\frac{1}{T - \Delta n}  \sum_{n=1}^{T- \Delta n}   \left( \tau_{n+ \Delta n} - {\overline \tau} \right) \left( \tau_{n} - {\overline \tau} \right)\ ,
\label{corr_iet}
\eeq
being $T$ the total number of IETs, $\Delta n$ the lag in the ``time'' index $n$ and:
$$
{\overline \tau} = \frac{1}{T} \sum_{n=1}^{T} \tau_n \ ;\quad \sigma^2_{\tau} = \frac{1}{T}  \sum_{n=1}^{T}   \left( \tau_{n} - {\overline \tau} \right)^2 
$$
the IET mean and variance. As usual, the variance is used as normalization factor, so that $C(0) = 1$.
It is important to underline that the autocorrelation function given in Eq. (\ref{corr_iet}) is not computed over time, but over the event occurrence index $n$, which moves from one event to the other, thus representing the ``internal'' time of the event sequence. From the usual time autocorrelation function it is derived the correlation time \cite{taylor_1921}, while a {\it correlation index} can similarly be computed from the above autocorrelation function:
%
\beq
\mathcal{T}_c = \sum_{\Delta n} C(\Delta n) \ .
\label{corr_time}
\eeq
where the zero time lag was omitted from the calculation because, by definition, it is constrained to the unitary value.
In fact, including $C(0)$ in the above sum, which is a discretized integral with unitary time step, would always add a unitary contribution that
does not provide accurate information about the time lag at which the system remains correlated. As a result, this could potentially level out the differences between the studied cases.
It is worth noting that the correlation time $\mathcal{T}_c$ was first introduced in diffusion studies \cite{taylor_1921,vitali_2018_hetero_fract} and that, in the case of exponential autocorrelation function $C(\Delta n) = exp(-\gamma \Delta n)$,  it is exactly given by the decay rate, i.e., $\mathcal{T}_c = 1/ \gamma$.
%

\subsubsection{Event definitions}
\label{event_def}

\noindent
In this work we considered two different kinds of crucial neural events: the {\it neural coincidence events} and the  {\it neural avalanche  birth-death events}, 
whose detailed definitions are given in Appendix \ref{app:coinc_avalanche}. 
In the case of avalanches, the IETs are exactly given by the avalanche duration times.
Avalanche is the basic self-organized structure introduced in Self-Organized Criticality (SOC) by Bak and co-workers \cite{bak_prl1987,bak_pra88} and later applied by Beggs and Plenz in their milestone papers discovering {\it neural avalanches} in {\it in vitro} neural networks \cite{beggsplenz_jn2003,beggsplenz_jn2004}.
Conversely, coincidence events were introduced in brain studies and applied to investigate the intermittency-based temporal complexity of electroencephalographic data \cite{fingelkurts_ijn2004,Kaplan2005RTPsAlgorithm,tozzi-fingelkurts_plr2017} (see \cite{paradisi_springer2017} for a review).
Coincidences are dynamical structures of the system that, by definition, are representative of a range of time scales shorter than those typical of avalanches, which are expected to evolve over longer time scales.
Consequently, IETs derived from coincidences are expected to be, on average, shorter in comparison to IETs based on avalanches.
%

\section{Results}
\label{results}

\subsection{Simulation setup}

\noindent
We carried out numerical simulations of the SEDAM model, as given in Eq. (\ref{our_model}), for different values of the parameters, namely the noise intensity $p$ and the number of patterns $K$.
The MNIST dataset \cite{deng2012mnist} was utilized to train the neural network. Specifically, comprehensive simulations were conducted by storing $K = 1, 10, 100, 1000$ and $10000$ binary images to assess how the network's dynamical behaviors evolve based on the number of stored patterns. 
To investigate the effects of noise and identify potential critical regions, $25$ distinct noise probabilities were evaluated: $p = 0.001, 0.01, 0.1, 0.2$, from $0.21$ to $0.4$ with incremental steps of $0.01$ and, finally,  $p=0.5$. It is worth noting that the values in the range $[0.5,1]$ have a symmetrical behavior with respect to the values in the range $[0,0.5]$, i.e., $p$ gives the same results as $1-p$.
The simulated networks consisted of $N=784$ neurons and were run for $T_{sim} = 200000$ time steps to ensure a thorough analysis of their temporal dynamics.
The initial condition of all numerical simulations is given by a distorted version of the pattern stored in the simulation with $K=1$.

\vspace{.2cm}
\noindent
In the following, we present our most important findings regarding the TC analyses applied to the SEDAM model, specifically providing a synthetic table of the statistical analyses.

\subsection{Temporal Complexity Analysis}

\begin{figure}[h!]
    \centering
    \includegraphics[scale=0.2]{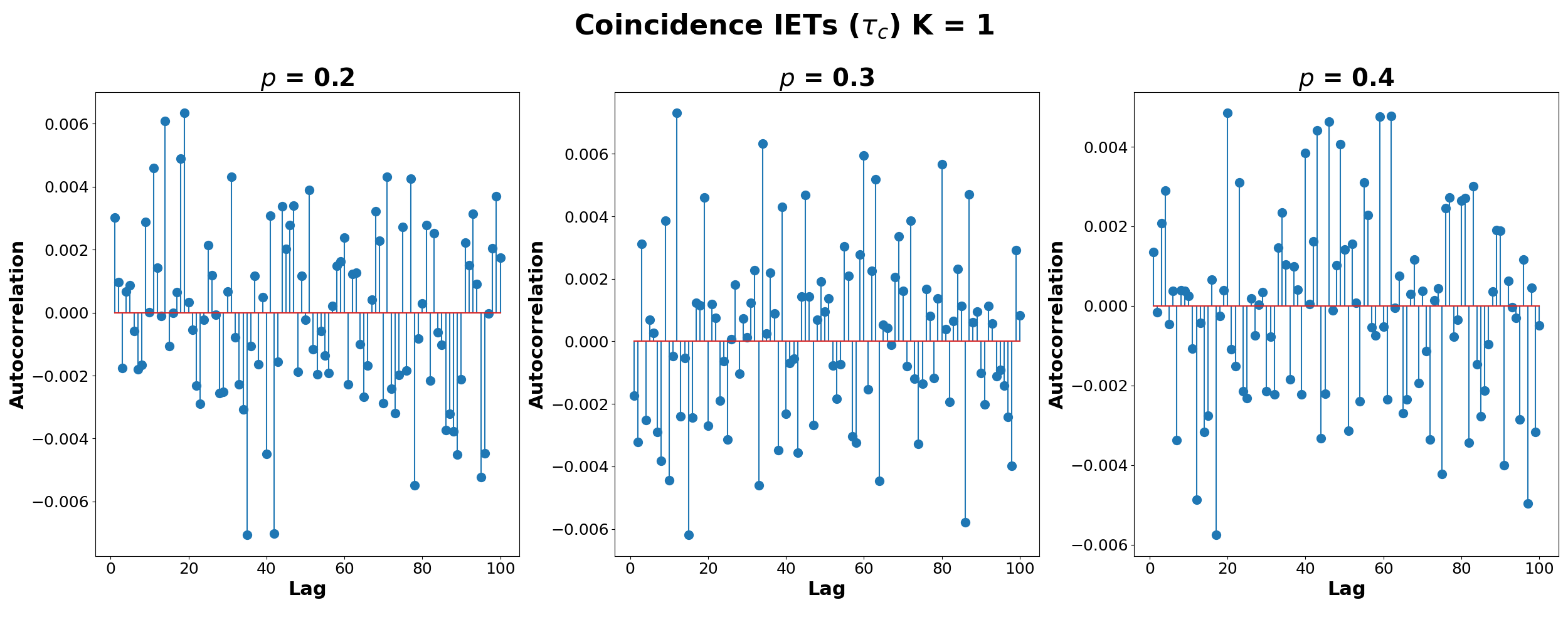}
    \includegraphics[scale=0.2]{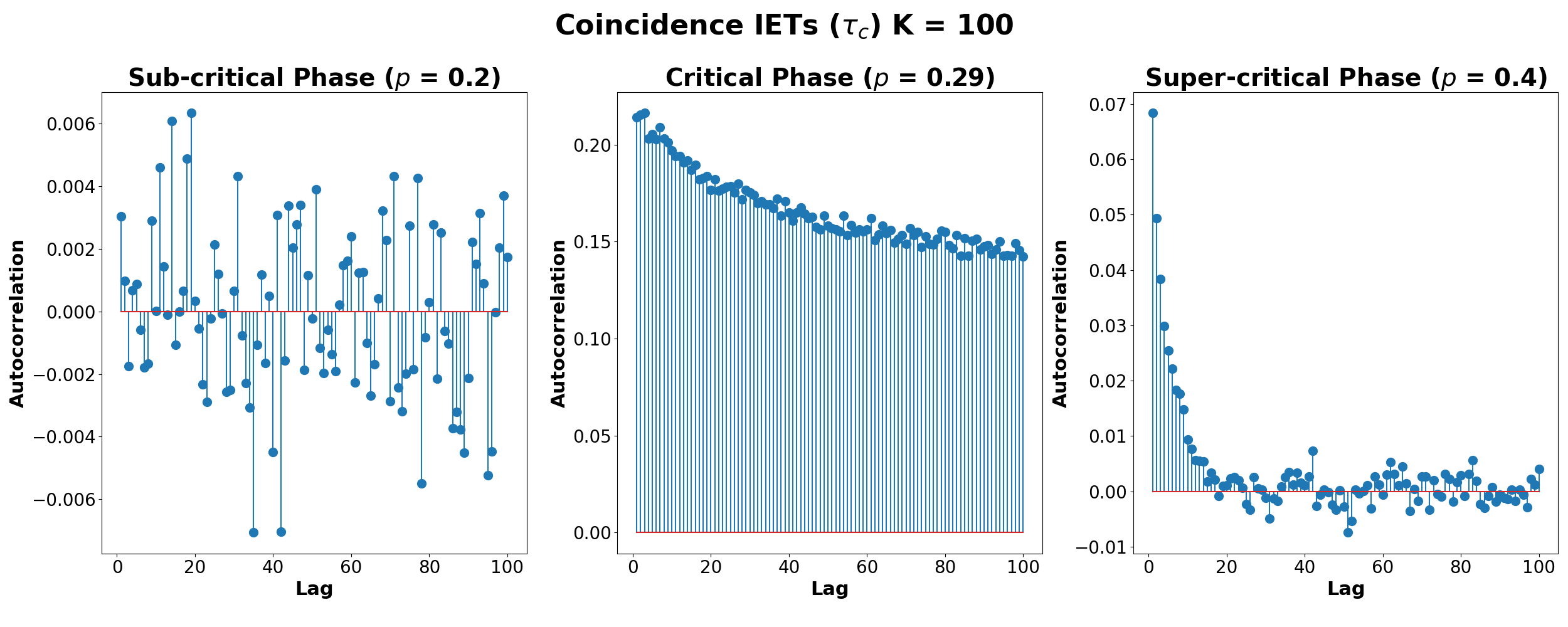}
    \caption{Coincidence events. IET autocorrelation functions with $K = 1$ (top panel) and $K = 100$ (bottom panel) stored patterns, shown across the three dynamical phases: sub-critical, critical, and super-critical (from left to right). ``Lag'' in the $x$-axis stands for $\Delta n$.}    \label{fig:Autocorrelation_Coincidences_IETs}
\end{figure}

\begin{figure}[h!]
\hspace{-3.cm}   
\def\svgwidth{19.cm}
\begingroup%
  \makeatletter%
  \providecommand\color[2][]{%
    \errmessage{(Inkscape) Color is used for the text in Inkscape, but the package 'color.sty' is not loaded}%
    \renewcommand\color[2][]{}%
  }%
  \providecommand\transparent[1]{%
    \errmessage{(Inkscape) Transparency is used (non-zero) for the text in Inkscape, but the package 'transparent.sty' is not loaded}%
    \renewcommand\transparent[1]{}%
  }%
  \providecommand\rotatebox[2]{#2}%
  \newcommand*\fsize{\dimexpr\f@size pt\relax}%
  \newcommand*\lineheight[1]{\fontsize{\fsize}{#1\fsize}\selectfont}%
  \ifx\svgwidth\undefined%
    \setlength{\unitlength}{576.09272634bp}%
    \ifx\svgscale\undefined%
      \relax%
    \else%
      \setlength{\unitlength}{\unitlength * \real{\svgscale}}%
    \fi%
  \else%
    \setlength{\unitlength}{\svgwidth}%
  \fi%
  \global\let\svgwidth\undefined%
  \global\let\svgscale\undefined%
  \makeatother%
  \begin{picture}(1,0.69754522)%
    \lineheight{1}%
    \setlength\tabcolsep{0pt}%
    \put(0,0){\includegraphics[width=\unitlength,page=1]{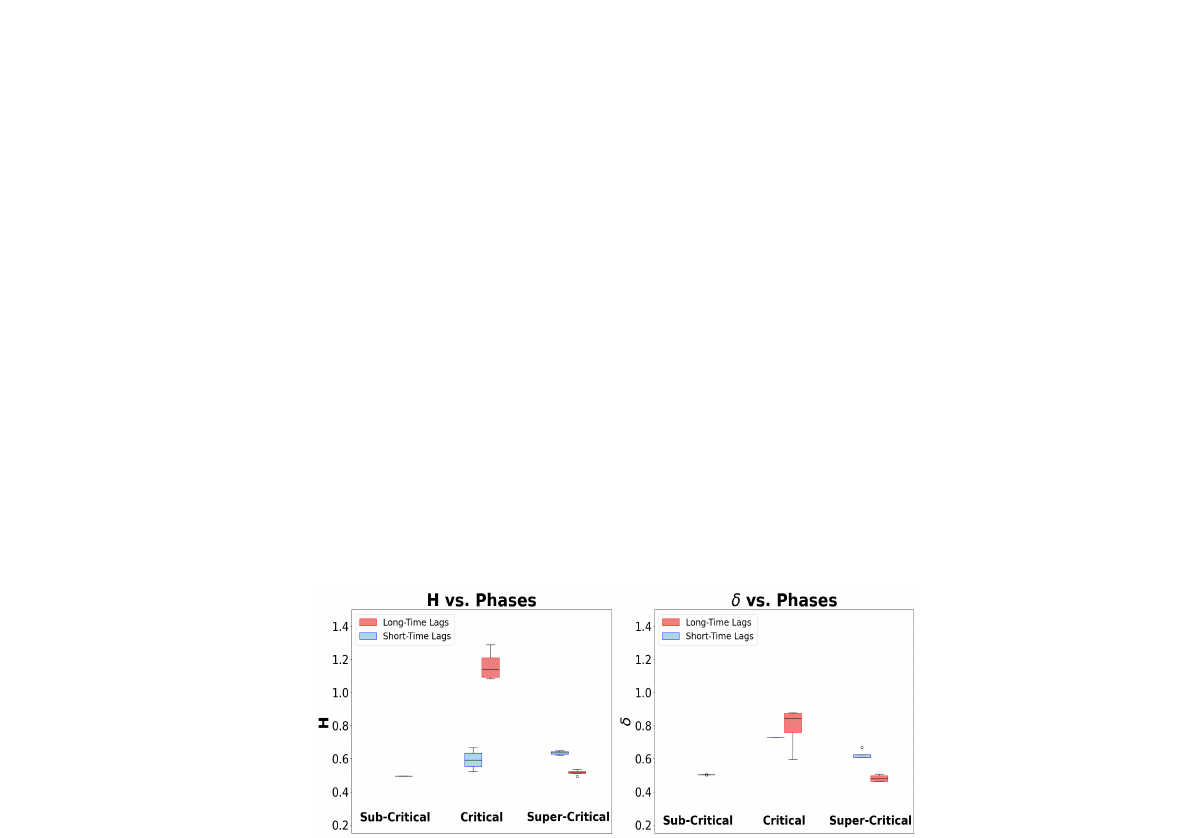}}%
    \put(0.26148695,0.19200112){\color[rgb]{0,0,0}\makebox(0,0)[t]{\lineheight{1.25}\smash{\begin{tabular}[t]{c}c)\end{tabular}}}}%
    \put(0,0){\includegraphics[width=\unitlength,page=2]{Coincidences_DFA_DE.pdf}}%
    \put(0.01718821,0.42825508){\color[rgb]{0,0,0}\makebox(0,0)[t]{\lineheight{1.25}\smash{\begin{tabular}[t]{c}b)\end{tabular}}}}%
    \put(0,0){\includegraphics[width=\unitlength,page=3]{Coincidences_DFA_DE.pdf}}%
    \put(0.0153863,0.67486138){\color[rgb]{0,0,0}\makebox(0,0)[t]{\lineheight{1.25}\smash{\begin{tabular}[t]{c}a)\end{tabular}}}}%
  \end{picture}%
\endgroup%

    \vspace{.1cm}
    \caption{Coincidence events. (a) DFA curves in the three different phases (sub-critical, critical, super-critical) for $K = 1, 10, 100, 1000$ stored patterns. (c) DE curves in the three different phases (sub-critical, critical, super-critical) for $K = 1, 10, 100, 1000$ stored patterns.
    Each panel legend shows the range of $p$ values relating to the phase indicated in the legend itself. 
    DFA and DE are referenced to the values $DFA(\Delta t_0)$ and $DE(\Delta t_0)$, respectively, being $\Delta t_0$ the smallest time lag at which DFA and DE are computed. The reported DFA and DE functions are derived as averages (points) and standard deviations (vertical bars) computed over the range of noise intensity values $p$ reported in the figure legend.
    (c) Boxplots of $H$ (left) and $\delta$  (right) values in the three phases (sub-critical, critical, super-critical) for both short- and long-time lag regimes. The crossover between short- and long-time ranges decreases with $K$. 
   }
\label{fig:H_delta_vs_phases_concidences}
\end{figure}

\begin{figure}[h!]
    \centering
    \includegraphics[scale=0.2]{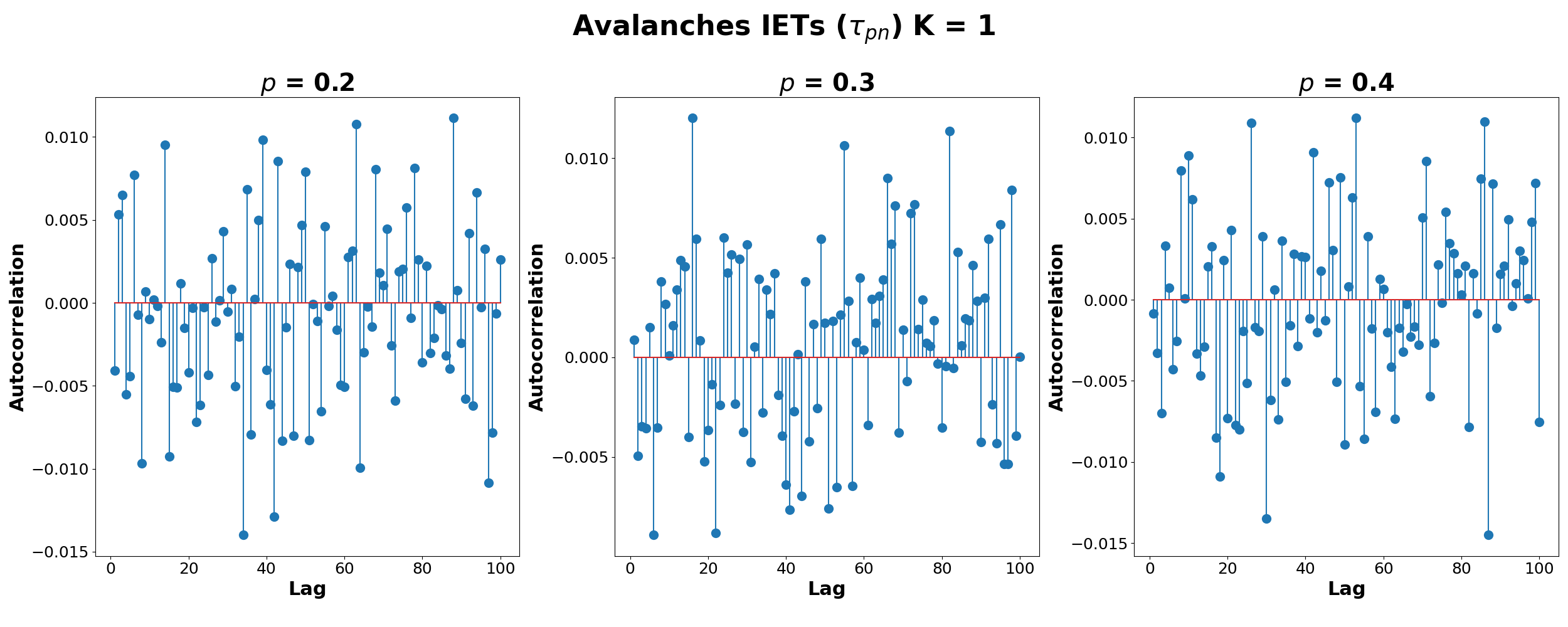}
    \includegraphics[scale=0.2]{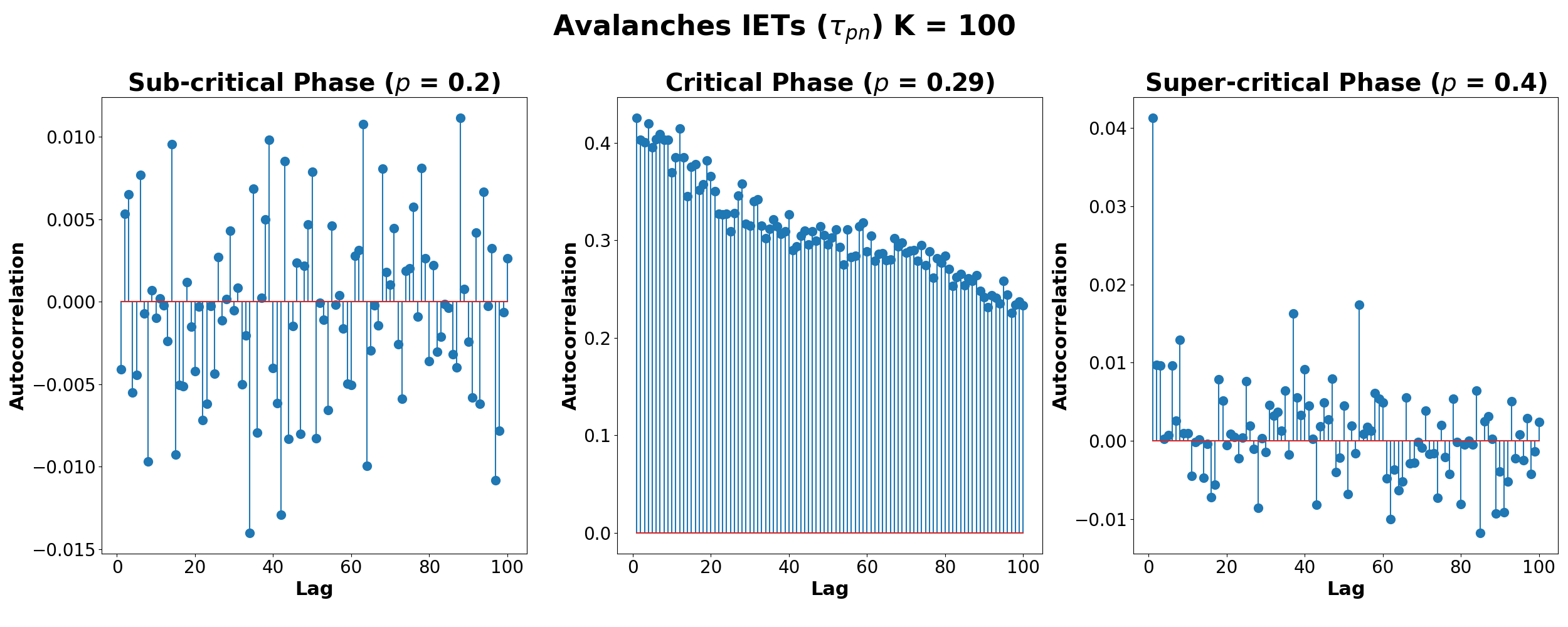}
    \caption{Avalanche birth-death events. IET autocorrelation function with $K = 1$ (top panel) and $K = 100$ (bottom panel) stored patterns, shown across the three dynamical phases: sub-critical, critical, and super-critical (from left to right). ``Lag'' in the $x$-axis stands for $\Delta n$.}
\label{fig:Autocorrelation_Avalanches_IETs}
\end{figure}

\begin{figure}[h!] 
\hspace{-3.cm}   
\def\svgwidth{19.cm}
\begingroup%
  \makeatletter%
  \providecommand\color[2][]{%
    \errmessage{(Inkscape) Color is used for the text in Inkscape, but the package 'color.sty' is not loaded}%
    \renewcommand\color[2][]{}%
  }%
  \providecommand\transparent[1]{%
    \errmessage{(Inkscape) Transparency is used (non-zero) for the text in Inkscape, but the package 'transparent.sty' is not loaded}%
    \renewcommand\transparent[1]{}%
  }%
  \providecommand\rotatebox[2]{#2}%
  \newcommand*\fsize{\dimexpr\f@size pt\relax}%
  \newcommand*\lineheight[1]{\fontsize{\fsize}{#1\fsize}\selectfont}%
  \ifx\svgwidth\undefined%
    \setlength{\unitlength}{576.09151525bp}%
    \ifx\svgscale\undefined%
      \relax%
    \else%
      \setlength{\unitlength}{\unitlength * \real{\svgscale}}%
    \fi%
  \else%
    \setlength{\unitlength}{\svgwidth}%
  \fi%
  \global\let\svgwidth\undefined%
  \global\let\svgscale\undefined%
  \makeatother%
  \begin{picture}(1,0.69754894)%
    \lineheight{1}%
    \setlength\tabcolsep{0pt}%
    \put(0,0){\includegraphics[width=\unitlength,page=1]{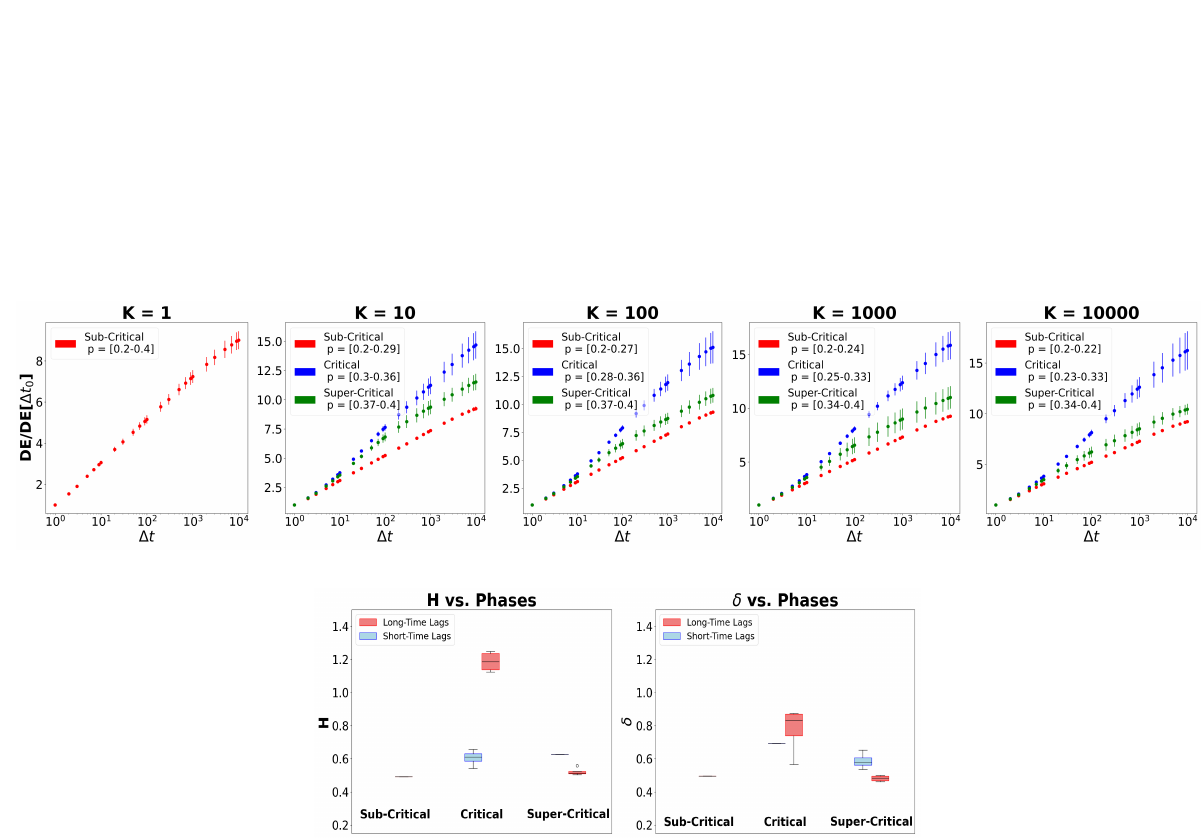}}%
    \put(0.01718821,0.42825508){\color[rgb]{0,0,0}\makebox(0,0)[t]{\lineheight{1.25}\smash{\begin{tabular}[t]{c}b)\end{tabular}}}}%
    \put(0.26148695,0.19200112){\color[rgb]{0,0,0}\makebox(0,0)[t]{\lineheight{1.25}\smash{\begin{tabular}[t]{c}c)\end{tabular}}}}%
    \put(0,0){\includegraphics[width=\unitlength,page=2]{Avalanches_DFA_DE.pdf}}%
    \put(0.0153863,0.67486138){\color[rgb]{0,0,0}\makebox(0,0)[t]{\lineheight{1.25}\smash{\begin{tabular}[t]{c}a)\end{tabular}}}}%
  \end{picture}%
\endgroup%

   \vspace{.1cm}
\caption{Same as Figure \ref{fig:H_delta_vs_phases_concidences}, but for avalanche birth-death events.}
    \label{fig:H_delta_vs_phases_avalanches}
\end{figure}

\noindent
%
After deriving the event sequences for both coincidences (IETs $\tau_c$) and avalanches (IETs $\tau_{pn}$),
IET autocorrelation, DFA and DE functions were computed.
Figures \ref{fig:Autocorrelation_Coincidences_IETs}-\ref{fig:H_delta_vs_phases_concidences} report the results regarding coincidence events, while Figures \ref{fig:Autocorrelation_Avalanches_IETs}-\ref{fig:H_delta_vs_phases_avalanches} display the results about avalanche birth-death events.
In the following we denote the {\it critical region} as the parametric region where an anomalous diffusion scaling ($H \ne 0.5$ and/or $\delta \ne 0.5$) and/or a correlation time $\mathcal{T}_c$ significantly different from zero are found.
Conversely, $H \simeq \delta \simeq 0.5$ and $\mathcal{T}_c \simeq 0$ identifies the sub-critical (low noise) and super-critical (high noise) regions.

%
%
%
%
%
%
%

\noindent
Figure \ref{fig:Autocorrelation_Coincidences_IETs} compares the coincidence events of the SEDAM models with $K=1$ (top panels) and $K=100$ (bottom panels). 
The $K=1$ model has no time correlation at all $p$ and, lacking any self-organizing capacity, can be depicted as a pure white noise.
Conversely, for $K=100$, an onset noise intensity value $p=p_c=0.29$ is observed, wherein the autocorrelation exhibits a significantly slower decay (bottom middle panel). 
The onset noise intensity value $p_c$ marks the transition to the critical region.\\
It is interesting to note that the sub-critical phase (bottom left panel) of the $K=100$ model is 
similar to the $K=1$ case (all top panels), while the super-critical phase (bottom right panel) is characterized by a correlation different from zero, which is however, still significantly lower than that of the critical phase.

\noindent
Similar behaviors are found for the autocorrelation of avalanche birth-death events, shown in figure \ref{fig:Autocorrelation_Avalanches_IETs}, but with stronger correlation at the onset of the critical region ($p=p_c=0.29$, bottom middle panel) and, surprisingly, a much lower avalanche event correlation than coincidence event correlation in the super-critical phase, at is can be depicted by comparing bottom right panels of Figures \ref{fig:Autocorrelation_Coincidences_IETs} and \ref{fig:Autocorrelation_Avalanches_IETs}. 

\noindent
Figure \ref{fig:H_delta_vs_phases_concidences}
gives a synthetic picture of the DFA and DE analyses applied to coincidence events. From the top panels, a), from left to right, it is possible to appreciate how the SEDAM model changes with increasing number of patterns $K$. Apart from the $K=1$ model, all models with $K>1$ display a well-defined critical region, which can be identified, in particular, by $H$ values that significantly depart from $0.5$, the reference value signaling the absence of long-range correlations.
As it is seen from middle panels b), the scaling $\delta$ of DE functions confirms the scaling $H$ of the DFA functions.
These findings are consistent with the results of the autocorrelation analysis. However, the DFA and DE functions provide a more pronounced display of the critical phase (blue dots/bars) and of the transition from the sub-critical phase (red dots/bars) to the super-critical phase (green dots/bars). 
In particular, in the DFA function of the critical region, it is possible to appreciate the decrease, as $K$ increases, of the time scale separating the normal scaling ($H=0.5$) at short times from the superdiffusive scaling  ($H>0.5$) at long times. This decrease, which ranges from about $\Delta t \approx 10^3$ ($K=10$) to $\Delta t \lesssim 10^2$ ($K=10000$), proves that the network's self-organizing capacity increases with $K$ in the critical phase, i.e., at proper, intermediate, noise intensities. On the contrary, the long-time scalings of the other phases remain compatible with normal diffusion ($H=0.5$), with the super-critical phase just showing a weak tendency to follow the same superdiffusive scaling of the critical phase in the short-time range. Notably, the short-time range of DE is more uniform across different $K$ values, being in the approximate interval $\Delta t \lesssim 10-20$.
In panel c), a synthetic view of the TC analyses is reported, showing the statistical boxplots of $H$ and $\delta$ evaluated with a fitting algorithm allowing to consider up to two different scalings at different ranges of $\Delta t$. Firstly, the boxplots of $H$ values (left panel) show a weak increase in the short-time range. However, the most interesting result is found in the long-time scaling, which reveals the net transition between the sub- and super-critical phases, both with normal diffusion ($H\simeq 0.5$), across the critical region, definitely giving values $H>1$ (average about $1.1$) and, thus, a strong superdiffusive regime. The DE scaling has a similar behavior in the long-time range, with an average $\delta \approx 0.85-0.9$ in the critical region. On the contrary, the short-time range has a slightly different behavior as it shows a transition even in this case, but a weaker one,  with the short-time scaling $\delta \lesssim 0.75$ in the critical region and $\delta \gtrsim   0.6$. However, it is worth noting that the accuracy of the short-time regime is lower than that of the long-time regime, as the shorter duration of the former affects its precision.

\noindent
The DFA and DE analyses applied to avalanche birth-death events, shown in Figure \ref{fig:H_delta_vs_phases_avalanches},
give patterns in the DFA and DE functions similar to those derived from coincidence events, with also similar crossover times between short- and long-time regimes. Panels c) show that only slight differences with respect to coincidence events are seen in the estimated values of $H$ and $\delta$. 
In more detail, 
the $H$ values (left panel c)) again show a weak increase in the short-time range, while the long-time scaling reveals a net transition between $H\simeq 0.5$ (sub- and super-critical phases) and $H\simeq 1.1-1.2$ (critical region). 
The avalanche-based DE scaling also shows behaviours, in both short- and long-time ranges, similar to the coincidence-based DE one, with slightly smaller values of $\delta$.

\noindent
In the Supplementary Material we also report some further qualitative analyses and discussions on the average network activity and on the IET-PDFs that may be useful in obtaining a more complete picture of phase transitions found through scaling analysis based on event-driven diffusion.

\subsection{Discussion}

\noindent
In Tables \ref{tab:coincidences_results_DFA_DE_Autocorr}
and \ref{tab:avalanches_results_DFA_DE_Autocorr}
we report, for the coincidence and avalanche events, respectively, the estimated values of $H$, $\delta$ and $\mathcal{T}_c$ in the three phases. Looking at these tables, it is evident the sharp transition in the diffusion scaling indices $H$ and $\delta$ and in the correlation index $\mathcal{T}_c$.
\begin{table}[h]
    \begin{tabular}{|c|c|c|c||c|c|c|c||c|c|c|}
        \hline
          & \multicolumn{3}{||c||}{  Sub-Critical ($p = 0.1$)}  & \multicolumn{4}{|c||}{Critical ($p = p_c$)} & \multicolumn{3}{|c|}{Super-Critical ($p = 0.4$)}\\
        \hline
         \multicolumn{1}{|c||}{$N$} & $H$ & $\delta$ & $\mathcal{T}_c$ & $p_c$ & $H$ & $\delta$ & $\mathcal{T}_c$ & $H$ & $\delta$ & $\mathcal{T}_c$ \\
        \hline
        \multicolumn{1}{|c||}{10} & 0.50 & 0.50 & -0.01 & 0.30 & [0.52, 1.29] & [0.73, 0.60] & 28.77 & [0.65, 0.49] & [0.67, 0.46] & 0.81 \\
        \hline
        \multicolumn{1}{|c||}{100} & 0.50 & 0.50 & -0.01 & 0.29 & [0.56, 1.09] & 0.87 & 16.58 & [0.62, 0.54] & [0.61, 0.46] & 0.35 \\
        \hline
        \multicolumn{1}{|c||}{1000} & 0.50 & 0.50 & -0.01 & 0.25 & [0.67, 1.18] & 0.81 & 22.17 & 0.52 & [0.61, 0.50] & 0.07 \\
        \hline
        \multicolumn{1}{|c||}{10000} & 0.50 & 0.50 & -0.01 & 0.23 & [0.62, 1.08] & 0.88 & 23.20 & 0.52 & [0.61, 0.51] & 0.07 \\
        \hline          
    \end{tabular}
    \caption{Coincidence events. $H$, $\delta$, and $\mathcal{T}_c$ in the sub-critical, critical and super-critical phases. The $p_c$ values reported in the table refer to the onset of the critical region starting from the sub-critical state.     
    $K=1$ is not reported as a critical state does not emerge in this case, i.e., $H = \delta \simeq 0.5$ and $\mathcal{T}_c \simeq 0$.}
\label{tab:coincidences_results_DFA_DE_Autocorr}
\end{table}
\begin{table}[h]
    \begin{tabular}{|c|c|c|c||c|c|c|c||c|c|c|}
     \hline
         & \multicolumn{3}{||c||}{Sub-Critical ($p = 0.1$)} & \multicolumn{4}{|c||}{Critical ($p = p_c$)} & \multicolumn{3}{|c|}{Super-Critical ($p = 0.4$)}\\
        \hline
        \multicolumn{1}{|c||}{$N$} & $H$ & $\delta$ & $\mathcal{T}_c$ & $p_c$ & $H$ & $\delta$ & $\mathcal{T}_c$ & $H$ & $\delta$ & $\mathcal{T}_c$ \\
        \hline
        \multicolumn{1}{|c||}{10} & 0.49 & 0.50 & -0.05 & 0.30 & [0.54, 1.25] & [0.69, 0.57] & 31.75 & [0.63, 0.51] & [0.65, 0.49] & 0.26 \\
        \hline
        \multicolumn{1}{|c||}{100} & 0.49 & 0.50 & -0.05 & 0.29 & [0.60, 1.12] & 0.87 & 30.94 & 0.56 & [0.59, 0.46] & 0.10\\
        \hline
        \multicolumn{1}{|c||}{1000} & 0.49 & 0.50 & -0.05 & 0.25 & [0.66, 1.23] & 0.80 & 35.89 & 0.51 & [0.53, 0.47] & 0.02 \\
        \hline
        \multicolumn{1}{|c||}{10000} & 0.49 & 0.50 & -0.05 & 0.23 & [0.62, 1.14] & 0.87 & 24.79 & 0.50 & [0.57, 0.50] & -0.01 \\
        \hline          
    \end{tabular}
    \caption{ Same as Table \ref{tab:coincidences_results_DFA_DE_Autocorr}, but for avalanche birth-death events.
    }    
\label{tab:avalanches_results_DFA_DE_Autocorr}
\end{table}
In more detail:
\begin{itemize}
    \item 
    The sub-critical phase always displays normal diffusion scaling $H\cong \delta \cong 0.5$ and $\mathcal{T}_c \simeq 0$; the retrieval dynamics are short-range correlated and Markovian. This corresponds to the well-known {\it retrieval phase}, in which the associative memory feature and recovery functioning have maximum efficiency, as the low noise intensity level does not allow to make jumps among energy wells. 
    %
    \item
    The super-critical phase displays a slightly greater variability in the short-time range, with estimated $H$ and $\delta$ values slightly above $0.5$ and, accordingly, non-zero values of $\mathcal{T}_c$  that, however, are found to be very small; conversely, the asymptotic long-time behavior falls back in the normal diffusion regime. In this phase, associative memory becomes ineffective and the network jumps from one energy well to another due to high noise intensities, independently from the initial pattern.   
    \item
    The critical region is not limited to a single value of $p$, but to an interval of values that, for each $K$, are reported in the legends of panels a) and b) in Figures \ref{fig:H_delta_vs_phases_concidences} and \ref{fig:H_delta_vs_phases_avalanches}. Within this range of $p$ values, both $H$ and $\delta$ exhibit a sharp deviation from the normal scaling $H=\delta=0.5$. Further, the noise intensity needed to maintain the critical phase decreases as $K$ increases. As expected, a smaller noise intensity level is sufficient in order to achieve, for a greater number of stored patterns, the critical and super-critical phases.  
    \item
    The differences in $H$ and $\delta$ among coincidence and avalanche events is minimal. A more significant difference is found in the correlation index $\mathcal{T}_c$, which is, in the critical phase, longer for avalanche than coincidence events  and, on the contrary, shorter in the super-critical phase. Thus, the differences in $\mathcal{T}_c$ affect the time scale at which superdiffusive scaling emerge, but not the scaling itself.
\end{itemize}
The values of $p_c$ reported in the tables refer to the onset value at which, starting from the sub-critical phase, a scaling $H$ and/or $\delta$ significantly different from $0.5$ emerges in the long-time regime. This implies that the reported $p_c$ values are, by definition, the minimal noise intensities needed to get strongly superdiffusive scaling indices $H$ and $\delta$ in the diffusion process generated by the detected events. As already said, this superdiffusive scaling is the signature of persistent long-range correlations, non-Gaussian and non-Markovian dynamics in the SEDAM model. 

\section{Concluding remarks}
\label{conclusions}

\noindent
In this paper, we have explored the temporal complexity (TC) of the SEDAM model, a stochastic version of the DAM with an exponential interaction function introduced by Demircigil et al. \cite{demircigil2017modernHopfExp}.
%
The pivotal discovery is the emergence of a critical region in the self-organizing features, specifically in the TC properties $H$, $\delta$ and $\mathcal{T}_c$, for noise intensities between a low-noise regime (sub-critical phase) and a high-noise regime (super-critical phase).
Even if the presence of a critical phase transition has already found in literature \cite{ramsauer2021hopfieldnetworksneed,lucibello_prl2024}, their findings are grounded on the {\it canonical ensemble} approach of equilibrium statistical mechanics, which cannot deal with the effective temporal evolution of the network dynamics. Conversely, here we do not assume any equilibrium hypothesis, and the SEDAM here studied is a dynamical system with multiplicative noise that, by greatly enriching the nonlinear network dynamics, has allowed us to study the TC features of coherent structures dynamically emerging in the network and, thus, its self-organizing capacity.
This self-organizing ability stems from the combined effect of the learning algorithm and noise.
It is worth noting that the authors of Ref. \cite{turalska_fp2012} discussed the potential application of TC to Hebbian learning.

\noindent
In more detail, the results of the analyses prove that the SEDAM reaches a maximum of temporal complexity in the critical region, while there is essentially no self-organizing in the other phases, apart from a very weak effect seen in the short-time range of the super-critical phase.
Interestingly, the temporal complexity of the critical region seems not to coincide with the canonical one that, in addition to non-Poisson
IETs, takes on the renewal assumption, that is, the mutual independence of IETs.
In fact, the concurrence of $H>1$ and of the large correlation index $\mathcal{T}_c$ effectively rules out the presence of renewal events. The SEDAM dynamics induced by the learning rule trigger a strong long-range correlation in both coincidence and avalanche events along their respective internal time indices. 
This condition of non-renewal ``temporal complexity" emerges at intermediate noise intensity values and characterizes the critical region 
(see Figures \ref{fig:H_delta_vs_phases_concidences} and \ref{fig:H_delta_vs_phases_avalanches}, panels a) or b)).
In agreement with the ``complexity'' concept, in the critical region the network works in between the ordered sub-critical phase and the disordered super-critical phase. Thus, in the critical phase the model preserves, at the same time, the recovery capacity, even if partially, but also maintain a good level of plasticity that allows to store other patterns.
Our results on the SEDAM model, which refers to the general framework of DAM models, are in qualitative agreement with those that Scarpetta and co-workers \cite{scarpetta2013NeuralAvalanchesatCriticalPoint} found in a leaky-integrate-and-fire (LIF) model with a spike-time-dependent-plasticity (STDP) mechanism for the learning phase, which is a temporal generalization of the Hebbian learning. In fact, they found that, in the critical regime, 
the retrieval of the two stored spatio-temporal patterns is not perfect and it is linked with the presence of power-law behaviors in the avalanche size and duration, while the retrieval greatly improves in the super-critical regime with higher excitability.
%
Interestingly, coincidence and avalanche events give very similar results regarding the diffusive scaling exponents $H$ and $\delta$. On the contrary, a significant difference lies in the correlation index $\mathcal{T}_c$, which is larger for avalanches than coincidences. This could appear to be an expected result, as avalanches evolve on longer times, but is actually not a trivial one, as the correlation index is based on the {\it internal} time of events, i.e., the event occurrence index $n$, and not on the observational time $t$.\\
As a final remark, we note that the critical behavior found in this work is not limited to a singularity point in the noise probability value $p$, but extends over substantial intervals that decrease slightly depending on the number of stored patterns $N$, but which overall fall within the range $p \approx 0.23 - 0.36$ (see panels (a) and (b) of Figs. \ref{fig:H_delta_vs_phases_concidences} and \ref{fig:H_delta_vs_phases_avalanches}).
This is then in agreement with the {\it extended criticality} concept introduced by Longo and co-workers \cite{bailly-longo_book2011,longo_fp2012} and, regarding temporal complexity, also found in a LIF neural network model \cite{lovecchio_fp2012}.

\bmhead{Supplementary information}
Supplementary Material can be found in \textcolor{red}{XXXXX}.

\bmhead{Acknowledgements}
This work was supported by the Next-Generation-EU programme under the funding schemes PNRR-PE-AI scheme (M4C2, investment 1.3, line on AI)
FAIR ``Future Artificial Intelligence Research'', grant id PE00000013, Spoke-8: Pervasive AI.

\section*{Statements and Declarations}

\begin{itemize}
\item {\bf Funding}\\ 
This work was supported by the Next-Generation-EU programme under the funding schemes PNRR-PE-AI scheme (M4C2, investment 1.3, line on AI)
FAIR ``Future Artificial Intelligence Research'', grant id PE00000013, Spoke-8: Pervasive AI.
\item {\bf Conflict of interest/Competing interests} \\
Both authors declare that they have no conflict of
interest.
%
%
\item {\bf Consent for publication}\\
Both authors approved the final version of the paper.
\item {\bf Data availability}\\
MNIST data were used for the numerical simulations and can be found on the public free repository (see Ref. \cite{deng2012mnist}).
%
%
\item {\bf Author contribution}\\
Both authors are main authors and contributed equally to the work with the same
roles, except: MC: Software, Visualization and PP: Supervision, Funding Acquisition, Project administration.
%
\end{itemize}

\appendix

\section{Event-driven diffusion scaling (EDDiS) Analysis}
\label{app:eddis}

\noindent
The Event-Driven Diffusion Scaling (EDDiS) algorithm, in its more general version, involves calculating three distinct random walks by applying different walking rules to a sequence ${t_n}$ of observed transition events
\cite{allegrini_pre2009,paradisi_romp12,paradisi_csf15_pandora,paradisi_springer2017}.
The scaling of these event-driven diffusion processes proved to be robust against the presence of noisy spurious events \cite{allegrini_pre10,paradisi_csf15_pandora,paradisi_springer2017} and, therefore, reliable in extracting the genuine complexity of the underlying complex event sequence. 
The EDDiS method includes computing the second moment scaling and the similarity of the diffusion probability distribution function (PDF). 
The underlying concept is rooted in the Continuous Time Random Walk (CTRW) model \cite{montroll1964random,weiss1983random}, where particle movement occurs only at event occurrence times.
Specifically, we focus on the Asymmetric Jump (AJ) walking rule \cite{grigolini2001asymmetric}, which involves making a single-step jump forward upon each event occurrence, effectively mirroring the counting process generated by the event sequence:
\beq
X(t) = \#\{n: t_n \le t\}.
\label{ctrw_aj}
\eeq

\noindent
For a renewal point process with a long-time power-law decay, i.e., $\psi(\tau) \sim 1/\tau^\mu$, exact relationships between $H$ ($\delta$) and the {\it temporal complexity index} $\mu$ are well known since decades (see \cite{paradisi_springer2017} for a brief survey).
In particular \cite{grigolini2001asymmetric}:
\begin{align}
\delta_{AJ}(\mu) &= 
\begin{cases}
    \mu - 1 & \quad 1 < \mu < 2 \\
    \frac{1}{\mu - 1} & \quad 2 \leq \mu < 3 \\
    \frac{1}{2} & \quad \mu \geq 3
\end{cases} \quad ; 
\hspace{.1cm}
&H_{AJ}(\mu) = 
\begin{cases}
    \frac{\mu}{2} & \quad 1 < \mu < 2 \\
    2 - \frac{\mu}{2} & \quad 2 \leq \mu < 3 \\
    \frac{1}{2} & \quad \mu \geq 3
\end{cases}
\label{aj_hdelta_vs_mu}
\end{align}
The value $H_{AJ}=\delta_{AJ}=0.5$  indicates Gaussian diffusion without memory, that is, with Markovian dynamics, so that the underlying sequence of events is generated by a non-cooperative process.
This condition is usually related to the onset of exponentially distributed IETs that are also statistically independent of each other, two conditions uniquely defining a renewal Poisson process.
Conversely, a non-Poisson process is associated to anomalous diffusion exponents, that is, $H_{AJ} \ne 0.5$ and/or $\delta_{AJ} \ne 0.5$. As said above and reported in Eq. (\ref{aj_hdelta_vs_mu}), renewal IETs with power-law decay regimes in their distribution, even if blurred with noisy spurious events\footnote{
Noisy events are those events generated by a random noise with short-range correlations and finite jumps and can be assumed to follow Poisson statistics \cite{allegrini_pre10,paradisi_csf15_pandora,paradisi_springer2017}.
},
cause a event-driven diffusion with well-defined scaling exponents $H$ and $\delta$ in the long-time regime. Interestingly, this has been observed also when the renewal condition is relaxed, even if a well-established theoretical framework does not still exist in this case and the relationship to the complexity index $\mu$ is not known.
In any case, the emergence of well-defined scaling exponents $H$ and/or $\delta$ is the signature of emerging self-organized states with
long-range (power-law) temporal correlations and non-Markovian dynamics. 

\subsection{Neural Coincidence events and Neural Avalanches}
\label{app:coinc_avalanche}

\noindent  
The TC features investigated here are applied to {\it neural coincidence events} and {\it neural avalanches}.

\noindent
Neural coincidence events are defined as instances where at least a minimum number $N_c$ of neurons fire simultaneously. Given a set of individual neuron activation times, the timing of a coincidence event is pinpointed as the instant when more than $N_c$ nodes activate within a brief window of duration $\Delta t_c$. In this study, we set $\Delta t_c$ to be equal to the simulation time step, i.e., $\Delta t_c = 1$, effectively meaning we are seeking perfectly concurrent events. Moreover, the $N_c$ value considered is equal to the $25^{th}$ percentile\footnote{
TC analyses based on $35^{th}$ percentile gave analogous results, in particular regarding the critical transition.
} 
of the coincidence size distribution computed for $\Delta t_c = 1$ and $N_c = 1$. 

\noindent
 Neural avalanches illustrate large-scale coordination among various brain regions and serve as a prime example of intermittent patterns in brain activity \cite{beggsplenz_jn2003, sorrentino_ni2023avalanchesbrainfingerprint}. Recent studies also have emphasized their importance in clinical contexts, revealing that disruptions in neural avalanche dynamics are associated with specific neurodegenerative disorders \cite{sorrentino_sc21avalanchesneurodegenerativedisease}. 
The definition of a neural avalanche begin by a firing counting carried out on the entire network over short time windows of length $\Delta t_{av}$. Then, a birth event time is the beginning of the first time windows when the firing number exceeds a given threshold $N_{av}$ after a previous time window with under-threshold counting.
Similarly, a death event time is the end of the last time windows when the
firing number exceeds the threshold $N_{av}$ before a next time window with under-threshold counting.
The avalanche duration time $\tau_{pn}$ is then computed as the difference between each death time and its immediately previous birth time.
During the avalanche the number of active neurons exceeds the threshold $N_{av}$ within each time window of length $\Delta t_{av}$ and remains sustained over time. Once avalanches are identified, their size is determined by counting the number of neuronal events occurring within each avalanche. 
Analogously to coincidences, we here set $\Delta t_{av} = 1$ and $N_{av}$ as the $25$-th percentile
of the coincidence size distribution, so that $N_{av} = N_c$.

\subsection{Detrended Fluctuation Analysis (DFA)}
\label{app:dfa}

\noindent
DFA is a well-known analysis (see, e.g., \cite{peng_pre94}) that is commonly employed in the scientific community for the evaluation of the second-moment scaling $H$ defined by:
\begin{eqnarray}
    &F^{^2}(\Delta t) = \langle \left( \Delta X(\Delta t) - \Delta X_{\rm trend}(\Delta t) \right )^2 \rangle \sim t^{2H}
    \label{h_scaling}\\
    \ \nonumber \\
    & F(\Delta t) = a \cdot \Delta t^H  \Rightarrow \nonumber \\
    & \Rightarrow  \log(F(\Delta t)) = \log(a)+ H \cdot \log(\Delta t) 
    \label{dfa_function}
\end{eqnarray}
being $\Delta X (t,\Delta t) = X(t+\Delta t) - X(t)$.
Scaling exponent is denoted as $H$ because it is essentially identical to the traditional Hurst similarity exponent, originally established by Hurst \cite{hurst_1951}. The expression $X_{\rm trend}(\Delta t)$ signifies an appropriate local trend within the time series. For a given time lag $\Delta t$, the DFA algorithm works as follows:
\begin{enumerate}
    \item 
    The time series is divided into $L$ non-overlapping time windows of duration $\Delta t$.
    \item
    For each time window $[t_l,t_l+\Delta t)$, $1\le l \le L$, a linear trend $X^l_{\rm trend}(t)$ is evaluated through a linear best fit in the time interval $t_l \le t < t_l+\Delta t$.  
    \item 
    In the $l$-th time window the second moment $\sigma^l(\Delta t)$ of the detrended signal $X(t) - X^l_{\rm trend}(t)$ is computed.
    \item 
    Finally, the DFA function $F(\Delta t)$ is defined by the square root of the average of $\sigma^l(\Delta t)$ over the $L$ time windows.
\end{enumerate}
The DFA is computed for different time lags $\Delta t$.
Within the EDDiS framework, the DFA is implemented on various event-driven diffusion processes, as elaborated in publications like \cite{paradisi_springer2017,paradisi_npg12}. To confirm the validity of Eq. \eqref{dfa_function} for the dataset and to determine the exponent $H$, one simply needs to conduct a linear regression in logarithmic coordinates.
For our DFA implementation, we employed the MFDFA function available in the MFDFA Python package \cite{gorjao2022MFDFA}.

\subsection{Diffusion Entropy (DE)}
\label{app:de}

\noindent
The DE analysis is used to estimate the probability self-similarity index $\delta$ \cite{akin_jsmte09,grigolini2001asymmetric}. In particular, it is defined as the Shannon Entropy of the diffusion process $X(t)$ and the following algorithm is used to estimate $\delta$:
\begin{enumerate}
      \item Time Series Partitioning: for a specified time lag $\Delta t$, separate the time series $X(t)$ into overlapping portions of length $\Delta t$. Then evaluate:
    $\Delta X (t, \Delta t) = X(t+\Delta t) - X(t),\ \forall\ t \in [0,t-\Delta t]$.
    \item Distribution Assessment: for each time lag \(\Delta t\), establish the distribution \(p(\Delta x,\Delta t)\).
    \item Shannon Entropy Computation: determine the Shannon entropy using the formula:
    \beq
    S(\Delta t) = - \int_{- \infty}^{+ \infty} p(\Delta x,\Delta t) \log p(\Delta x,\Delta t) dx
    \label{de_shannon}
    \eeq
    When the probability density function (PDF) is self-similar, i.e., \(p(\Delta x,\Delta t) = f(\Delta x / \Delta t^\delta)/\Delta t^\delta\), then:
    \beq
    S(\Delta t) = A + \delta \log (\Delta t + T)
    \label{de_scaling}
    \eeq
    To verify the accuracy of Eq. \eqref{de_scaling} on the data and to estimate the exponent $\delta$, it is sufficient to perform a linear best fit on a lin-log scale, with logarithmic scale along the temporal axis.
\end{enumerate}

\bibliography{Bibliography}

\end{document}